\documentclass[preprint,superscriptaddress,amsmath,amssymb]{revtex4-1}

\usepackage{times}
\usepackage{graphicx}
\usepackage{dcolumn}
\usepackage{amsmath}
\usepackage{color}
\usepackage{caption,setspace}
\begin{document}

\title{Dielectric Metasurfaces for Complete Control of Phase and Polarization with Subwavelength Spatial Resolution and High Transmission}

\author{Amir Arbabi}
\affiliation{T. J. Watson Laboratory of Applied Physics, California Institute of Technology, 1200 E California Blvd., Pasadena, CA 91125, USA}
\author{Yu Horie}
\affiliation{T. J. Watson Laboratory of Applied Physics, California Institute of Technology, 1200 E California Blvd., Pasadena, CA 91125, USA}
\author{Mahmood Bagheri}
\affiliation{Jet Propulsion Laboratory, California Institute of Technology, Pasadena, CA 91109, USA}
\author{Andrei Faraon}
\affiliation{T. J. Watson Laboratory of Applied Physics, California Institute of Technology, 1200 E California Blvd., Pasadena, CA 91125, USA}

\maketitle

\textbf{Metasurfaces are planar structures that locally modify the polarization, phase, and amplitude of light in reflection or transmission, thus enabling lithographically patterned flat optical components with functionalities controlled by design~\cite{Kildishev2013a,Yu2014}. Transmissive metasurfaces are especially important, as most optical systems used in practice operate in transmission. Several types of transmissive metasurfaces have been realized ~\cite{Yu2011,Lin2014a,Lin2013,Vo2014}, but with either low transmission efficiencies or limited control over polarization and phase. Here we show a metasurface platform based on high-contrast dielectric elliptical nano-posts which provides complete control of polarization and phase with sub-wavelength spatial resolution and experimentally measured efficiency ranging from 72\% to 97\%, depending on the exact design. Such complete control enables the realization of most free-space transmissive optical elements such as lenses, phase-plates, wave-plates, polarizers, beam-splitters, as well as polarization switchable phase holograms and arbitrary vector beam generators using the same metamaterial platform.}

Polarization, phase, and amplitude completely characterize monochromatic light. In free space optical systems, polarization is modified using wave retarders, polarizers, and polarization beam splitters, phase is shaped using lenses, curved mirrors or spatial phase modulators, and amplitude is controlled via neutral density absorptive or reflective filters. Several metasurface platforms have been recently investigated to replicate the functionality of common optical components such as wave retarders, polarizers, beam splitters, lenses, or focusing mirrors. However, none of the platforms achieves complete control of both polarization and phase with sub-wavelength sampling and high transmission. A metasurface platform achieves complete control over polarization and phase if it can generate any desired physically admissible spatially varying polarization and phase distributions from an input beam with arbitrary polarization and phase distributions. Plasmonic metasurfaces have been demonstrated, but they have limited efficiencies because of fundamental limits \cite{Monticone2013, Arbabi2014b} and metal absorption loss \cite{Aieta2012,Pfeiffer2013,Lin2013}. Components based on 1D high contrast gratings have higher efficiencies, but do not provide high spatial resolution for realizing precise phase or polarization profiles in the direction along the grating lines~\cite{Fattal2010,Lu2010,Klemm2013,Lin2014a,Aieta2015}. The majority of flat elements have been realized using a platform that provides only phase control \cite{Warren1995,Lalanne1999,Fattal2010,Yu2011,Klemm2013,Lin2014a,Vo2014,Arbabi2014,Arbabi2015,Vo2014,West2014,Decker2015} (in most cases only for a fixed input polarization), or only a limited polarization modification capability  \cite{Kikuta1997,Lin2013,Schonbrun2011,Yang2014,Mutlu2012c,Zhao2012}. The platform we propose does not suffer from these limitations and provides a unified framework for realizing any device for polarization and phase control with an average transmission higher than 85\%.

A generic illustration of a transmissive metasurface which provides spatially varying control of phase and polarization for monochromatic light is shown in Fig. 1a. In this illustration, the metasurface is divided into hexagonal pixels, but other lattice types could also be chosen. An optical wave with spatially varying electric field $\mathbf{E}^\mathrm{in}$ is incident on the metasurface. The polarization ellipse and the phase of the optical field $\mathbf{E}^\mathrm{out}$ transmitted through each pixel can be controlled arbitrarily via the pixel design (Fig. 1a, top view). To avoid diffraction of light into non-zero diffraction orders, and to achieve high polarization and phase gradients required for implementation of optical components such as lenses with high numerical apertures,  it is important that each pixel has a lateral dimension smaller than a wavelength. The general relation between the electric fields of the input and output waves at each pixel is expressed using the Jones matrix ($\mathbf{T}=\bigl[\begin{smallmatrix}T_{xx}&T_{xy}\\ T_{yx}&T_{yy}\end{smallmatrix} \bigr]$) of the pixel as $\mathbf{E}^\mathrm{out}=\mathbf{T}\mathbf{E}^\mathrm{in}$. It is shown in the Supplementary Section S1 that, for metasurfaces with high transmission, any arbitrary ($\mathbf{E}^\mathrm{in}$) can be mapped to any desired ($\mathbf{E}^\mathrm{out}$) using a symmetric and unitary Jones matrix. Therefore, a metasurface platform achieves complete polarization and phase control if it can implement any unitary and symmetric Jones matrix at each pixel.

Figure 1b shows a schematic illustration of the proposed metasurface platform. It is composed of a single layer array of amorphous silicon elliptical posts with different sizes and orientations, resting on a fused silica substrate. The posts are placed at the centres of hexagonal unit cells. In a simplified picture, each post can be considered a waveguide which is truncated on both sides and operates as a low quality factor Fabry-P\'{e}rot resonator. The elliptical cross section of the waveguide leads to different effective refractive indices of the waveguide modes polarized along the two ellipse diameters. As a result, each of the posts imposes a polarization dependent phase shift to the transmitted light and modifies both its phase and polarization. In the regime of operation considered here (also discussed previously in~\cite{Arbabi2015}) light is mainly confined inside the high refractive index posts which behave as weakly coupled low quality factor resonators. Therefore, the light scattered by each post is primarily affected by the geometrical parameters of the post and has negligible dependence on the dimensions and orientations of its neighbouring posts.  As a result, each unit cell of the lattice can be considered as a pixel similar to the illustration shown in Fig. 1a.

Light scattering by high refractive index single dielectric scatterers has been studied previously, and it has been shown that they may possess strong effective magnetic dipoles and exhibit large forward scattering~\cite{Garcia-Etxarri2011, Evlyukhin2011, Spinelli2012} (see Supplementary Fig. 1). Here, instead of studying the properties of a single elliptical post, we use a different approach and examine the transmission properties of periodic arrays of weakly coupled posts. Such periodic arrays better approximate the local transmission properties of a metasurface composed of gradually varying posts. We use the Jones matrix of the periodic arrays to approximate the local Jones matrix of each pixel. This approximation is used here to successfully realize high performance devices for polarization and phase control (discussed later in Fig. 4 and Fig. 5), thus further validating its accuracy. 

A periodic array of elliptical posts with one ellipse axis aligned to one of the hexagonal lattice vectors ($\mathbf{a}_1$ which is along the $x$ axis) is shown in Fig. 2a. Due to symmetry, a normally incident optical wave linearly polarized along one of the ellipse axes does not change polarization and only acquires phase as it passes though the array. The phase shifts imposed by the array to the $x$ and $y$-polarized waves (i.e. $\phi_x$ and $\phi_y$) are functions of the elliptical post diameters $D_x$ and $D_y$. Therefore, the array behaves as a 2D material with adjustable birefringence whose principal axes are along $x$ and $y$ directions. The phases ($\phi_x$ and $\phi_y$) and the  intensity transmission coefficients ($|t_x|^2$ and $|t_y|^2$) were first determined via simulations as functions of the ellipse diameters (see Methods and Supplementary Figure 2 for details). From these simulations, the required diameters $D_x$ and $D_y$ to achieve all combinations of $\phi_x$ and $\phi_y$ while maintaining high transmission were derived, and are plotted in Fig. 2b and 2c. Any combination of $\phi_x$ and $\phi_y$ can be simultaneously obtained by properly choosing $D_x$ (from Fig. 2b) and $D_y$ (Fig. 2c).  The corresponding intensity transmission coefficients (i.e. $|t_x|^2$ and $|t_y|^2$) are shown in Fig. 2d and 2e, and are larger than 87\% for all values of $\phi_x$ and $\phi_y$. The complete phase coverage combined with the high transmission result in the high performance of this platform.

The principal axes of the birefringent array shown in Fig. 2a can be rotated by rotating the entire array or, with good approximation, by rotating all posts around their axes. This can be seen in Fig. 3 which shows that rotating the posts around their axes leads to approximately the same Jones matrix elements as rotating the entire array by the same angle. This is a result of the confinement of the optical energy inside the posts (as can be seen in Fig. 3b) which has led to the weak coupling among the posts~\cite{Arbabi2015}. This also provides another evidence that the polarization and phase transformation by the elliptical posts can be considered as a local effect.

As we mentioned earlier, a metasurface can achieve complete polarization and phase control if each of its pixels can be designed to realize any unitary and symmetric Jones matrix. It is shown in the Supplementary Section S2 that any desired symmetric and unitary Jones matrix can be realized using a birefringent metasurface if $\phi_x$, $\phi_y$, and the angle between one of the principal axes of the material and the $x$ axis ($\theta$) could be chosen freely. As Figs. 2 and 3 show, all these degrees of freedom are achievable at each pixel through selection of the post diameters $D_x$ and $D_y$, and its in-plane rotation angle $\theta$ (see the inset of Fig. 1b). Therefore, any desired spatially varying polarization and phase profiles can be generated by sampling the incident wavefront with a sub-wavelength lattice, and placing elliptical posts with proper dimensions and rotation angles at the lattice sites to impart the required phase and polarization change on the transmitted light. The proposed metasurface platform samples both the incident and the transmitted wavefront with sub-wavelength resolution in both silica and air because the reciprocal vectors of the lattice are larger than the wavenumber of light and thus, for close to normal incidence, the first order diffraction is not present.

The freedom provided by the proposed platform to simultaneously control the polarization and phase of light allows for implementation of a wide variety of optical components. To demonstrate the versatility and high performance of this platform, we fabricated and characterized two categories of flat optical elements operating at the near infrared wavelength of 915 nm. The devices consist of 715 nm tall amorphous silicon posts with diameters ranging from 65 nm to 455 nm, arranged on a hexagonal lattice with 650 nm lattice constant (for details of fabrication and measurement, see Methods). The first category of devices generate two different wavefronts for two orthogonal input polarizations. This functionality can be achieved if the device does not change the polarization ellipses of the two orthogonal polarizations it has been designed for, and only changes their handedness (see Supplementary Section S3). A special case is when both of the input polarizations are linear. Simulation and experimental measurement results, as well as the optical and scanning electron microscope images of three types of devices in this category are shown in Fig. 4. A polarization beam splitter that deflects $x$ and $y$ polarized portions of light by 5$^\circ$ and -5$^\circ$ is presented in Fig. 4a. We measured 72\% and 77\% efficiencies for the $x$ and $y$-polarized input light, respectively. The measured efficiencies are smaller than their corresponding simulated values (89\% for $x$ and 93\% for $y$-polarized incident beams) because of slight differences between the diameters of the designed and fabricated posts. A polarization beam splitter which separates and focuses the $x$ and $y$ polarized light at two different positions is presented in Fig. 4b. The focusing efficiency (defined as the ratio of the optical power focused to the desired spot to the input power) was measured as 80\% and 83\% for the $x$ and $y$-polarized  light, respectively. A polarization switchable phase hologram which generates two distinct patterns for $x$ and $y$-polarized light is shown in Fig. 4c. The change of the recorded pattern with polarization is shown in Supplementary Video 1. This is the most general form of device from this category.  We measured efficiencies of 84\% and 91\% for this device for the $x$ and $y$-polarized incident light. The measured intensity profiles presented in Fig. 4 show the total transmitted light as detected by the camera and there is no background subtraction. 

The second category of devices generate light with a desired arbitrary phase and polarization distribution from an incident light with a given polarization. Fig. 5a shows a device that transforms an incident $x$-polarized incident Gaussian beam into a radially polarized Bessel-Gauss beam, and a $y$-polarized incident Gaussian beam into an azimuthally polarized Bessel-Gauss beam.  We measured a transmission efficiency of 96\% and 97\% for the $x$ and $y$ input polarizations, respectively. Measured intensity profiles for different polarization projections are also shown in Fig. 5. When the polarization of the incident Gaussian beam is linear but not aligned with the $x$ or $y$ axis, a generalized cylindrical vector beam is generated by this device. It has been recently shown that cylindrical vector beams show unique features such as focus shaping when focused with a high numerical aperture lens~\cite{Zhan2009}. Furthermore, the same device in Fig. 5a generates light with different orbital angular momentum depending on the helicity of the input beam; right and left handed circularly polarized input beams will respectively acquire $m=1$ and $m=-1$ units of orbital angular momentum as they pass through this device. Both the generation and focusing of cylindrical vector beams can be performed using a single device based on the proposed platform. Such a device which simultaneously generates and focuses radially and azimuthally polarized light is shown in Fig. 5b. Similar to the device shown in Fig. 5a, due to the polarization conversion, right and left handed polarized beams acquire plus or minus one units of orbital angular momentum as they pass through the device. As a result, by adding a sinusoidal dependence in the form of $\mathrm{exp}(i\phi)$ to the phase profile of the device, the total orbital angular momentum of the right and left handed circularly polarized light after passing through the device will become $m=0$ and $m=2$, respectively. A device with such a phase and polarization profile is shown in Fig. 5c. As can be seen from the simulation and measurement results, a right handed circularly polarized incident beam is focused to a nearly diffraction limited spot (see Supplementary Fig. 3) while a left handed circularly polarized beam is focused into a doughnut shaped intensity pattern. Therefore, the focal spot shape can be modified by changing the polarization of the incident beam. This is particularly interesting since the polarization state of the incident beam can be switched rapidly using a phase modulator. 
 
 The functionalities provided by some of the optical devices demonstrated here can only be achieved by using a combination of multiple bulk optical components. For example, to realize the functionality of the polarization beam splitter and focuser in Fig. 4b, a Wollaston prism and two carefully aligned lenses are required. The realization of the polarization vector beams shown in Fig. 5 generally requires interferometry \cite{Phelan2011}, liquid crystal spatial light modulators, or conical Brewster prisms \cite{Kozawa2005}. The complete and simultaneous control over the polarization and phase profiles of light offered by the proposed platform and the design technique enables realization of novel optical components with functionalities exactly tailored to specific applications and with form factors required for emerging applications like wearable consumer electronics. As most other diffractive optical elements, these devices have an optical bandwidth of several percent of the design wavelength \cite{Swanson1989}. Thus, they could directly replace conventional optics in applications employing narrow-band light sources such as optical communications, monochromatic imaging and multi-photon microscopy.  We note that the theoretical approach and the design technique we introduced here are general and applicable to similar platforms with other types of scatterers and lattice shapes provided they offer complete and independent phase control for two orthogonal polarizations. The operating wavelength can also be changed by scaling the device dimensions. Further improvements are expected from using materials with optical nonlinearities and gain that might extend the spectral bandwidth of operation and provide tunability. These metasurfaces could also be patterned on curved or flexible substrates thus enabling conformal optical devices.

\clearpage

\section*{Methods}
\textbf{Simulations and design.} To obtain the simulation results presented in Figs. 2b and 2c, we computed the transmission coefficients $t_x$ and $t_y$ of $x$ and $y$-polarized plane waves for the periodic array shown schematically in Fig. 2a by using the rigorous coupled wave analysis (RCWA) technique using a freely available software package~\cite{Liu2012}. The simulations were performed at $\lambda=$915~nm. The amorphous silicon posts (with refractive index of 3.56 at 915~nm) are 715~nm tall, and rest on a fused silica substrate. We computed these transmission coefficients for all mutual values of the ellipse diameters $D_x$ and $D_y$ in the range $0.1a$ to $0.7a$, where $a=$650 nm is the lattice constant.  Simulation results are presented in Supplementary Fig. 2. For normal incidence, the array is non-diffractive in both air and fused silica at wavelengths longer than $\lambda_1=n_\mathrm{SiO_2}\sqrt{3}/2a=$816 nm. Next, using the computed transmission coefficients, for all combinations of the phases $\phi_x$ and $\phi_y$ we found the diameters $D_x$ and $D_y$ that minimizes the mean squared error $E=1/2(|t_x-\mathrm{e}^{i\phi_x}|^2+|t_y-\mathrm{e}^{i\phi_y}|^2)$. It should be noted that the elliptical posts  obtained using this method do not possess resonances close to the operation wavelength since the transmission values go to zero at resonance and increases the mean squared error. Transmission spectra for a few arrays with different diameters are presented in Supplementary Fig. 4. 

The simulation results presented in Figs. 3b and 3c were also computed using the RCWA technique. In Fig. 3b, an $x$-polarized light with magnetic energy density of 1 is normally incident on the posts from the top. The simulation parameters used to obtain Figs. 3b and 3c are the same as the ones used in Fig. 2, and the diameters of the elliptical posts are 300 nm and 150 nm.

  To design the devices presented in Fig. 4 which impose two distinct phase profiles to $x$ and $y$ polarized light, the optimum phase profiles that generate the desired patterns were first determined by back propagating the desired pattern to the plane of the device and finding the phase difference between the back propagated wave and the incident wave. This method is discussed in details in \cite{Arbabi2015}. After finding the desired phase profiles for both of the polarizations, the profiles were sampled at the lattice sites and elliptical posts with major and minor diameters that impart the required phases and polarization rotations on the transmitted beam were placed on those sites.
  
Optical elements shown in Fig. 5 that simultaneously modify polarization and phase of light were designed to generate a desired spatially varying optical wave from a given input optical wave profile. We first sampled the input and output optical waves at the lattice sites. Next, we computed the Jones matrix and decomposed it into its eigenvalues and eigenvectors to determine the desired phase shifts for waves polarized along the axes of the ellipse (i.e. $\phi_x$ and $\phi_y$) and the rotation angle $\theta$ (see Supplementary Sections S1 and S2 for details). Finally, from Figs. 2b and 2c, we found the diameters of the elliptical posts imposing the target $\phi_x$ and $\phi_y$ phase shifts, rotated them anticlockwise by their $\theta$, and placed them at their corresponding lattice sites.

We computed the simulation results presented in Figs. 4 and 5 by assuming that the devices perform the polarization and phase conversions ideally, and with $\lambda/15$ spatial resolution. For these simulations, the input light was assumed to be uniformly polarized Gaussian beams with the same beam radius as the illumination beam radius used in the corresponding measurements (50 $\mu$m for the device shown in Fig. 4a and 80 $\mu$m for the devices shown in Figs. 4b,c and 5). The output light was computed at each point on a rectangular grid assuming ideal polarization and phase conversion by the device, and then propagated to the planes of interest using the plane wave expansion technique \cite{Born1999}.  

\textbf{Sample fabrication.} We fabricated the devices shown in Figs. 4 and 5 on a fused silica substrate. We deposited 715 nm hydrogenated amorphous silicon using plasma enhanced chemical vapour deposition (PECVD) with a 5\% mixture of silane in argon at 200$^\circ$ C. Next, we spun 300 nm positive electron beam resist (ZEP-520A) and $\sim$60 nm of a water-soluble anti-charging conductive polymer (aquaSave from Mitsubishi Rayon) to avoid static charging during electron beam lithography. The pattern was written on the resist using electron beam lithography, the anti-charging layer was removed in water, and the pattern was developed in a resist developer (ZED-N50 from Zeon Chemicals). A 70 nm thick aluminium oxide layer was then deposited on the developed resist and patterned by lifting off the resist. The patterned aluminium oxide was subsequently used as a hard mask for dry etching of amorphous silicon in a 3:1 mixture of $\mathrm{SF_6}$ and $\mathrm{C_4F_8}$. Finally, the aluminium oxide mask was removed using a 1:1 mixture of ammonium hydroxide and hydrogen peroxide heated to 80$^\circ$ C.

To compensate for the systematic fabrication errors such as non-optimum exposure dose in the e-beam lithography patterning and a possible undercutting during dry etching, for each of the devices shown in Figs. 4 and 5, we fabricated a series of devices with all post diameters uniformly biased from their optimum design values by steps of 5 nm. Characterization results for the devices with different diameters showed that the device functionalities were not severely affected by these intentionally introduced systematic errors and only the device efficiency was reduced from their maximum value by approximately 3\% per each 5 nm error in the post diameters.

\textbf{Measurement procedure.} We characterized the devices using a setup that is schematically shown in Supplementary Fig. 5a. Light from a 915 nm fibre coupled semiconductor laser was passed through a fibre polarization controller and collimated to generate a Gaussian beam. To collimate the fibre output and generate a Gaussian beams with a beam radius smaller than the device radius, we used a fibre collimation package (Thorlabs F220APC-780) and a lens (Thorlabs LB1676-B with focal length of 10 cm for measuring the devices shown in Fig. 4a and Thorlabs LB1945-B with focal length of 20 cm for the devices shown in Figs. 4b,c and 5). The illumination beam radius on the sample was adjusted by changing the distance between the lens and the sample. The beam radius was set to approximately 35 $\mu$m for measuring the device shown in Fig. 4a to avoid overlap of the deflected and non-deflected portions of the output light at the measurement plane. To fill most of the device physical aperture, the illumination beam radius was set to 80 $\mu$m for all other measurements reported in Figs. 4b,c and 5.
 The objective lens, the tube lens (Thorlabs LB1945-B), and the camera (CoolSNAP K4 from Photometrics) shown in Supplementary Fig. 5a comprise a custom built microscope. We used three different objective lenses to achieve different magnifications. Measurement results shown in Fig. 4a were obtained using a 20X objective lens (Olympus UMPlanFl, NA=0.4), results shown in Figs. 4b and 4c, and Fig. 5a were recorded using a 50X objective lens (Olympus LCPlan N, NA=0.65), and the ones presented in Figs. 5b and 5c were obtained using a 100X objective lens (Olympus UMPlanFl, NA=0.95). The overall microscope magnification for each objective lens was found by imaging a calibration sample with known feature sizes. The polarizer (Thorlabs LPNIR050-MP) was inserted into the setup to confirm the polarization state of the incident light (after removing the device) and the output light. Efficiency values for the devices shown in Figs. 4a,c and 5 were obtained by integrating the light intensity on the camera (i.e. the measured intensity profiles shown in Figs. 4 and 5), subtracting the dark noise, and normalizing it to the integrated intensity recorded when the device was removed. For the device shown in Fig. 4a, only the intensity of the deflected portion of the output light was used for efficiency calculation. To characterize the efficiency of the device shown in Fig. 4b, we used a setup schematically shown in Supplementary Fig. 5b. A 25 $\mu$m diameter pinhole (Thorlabs P25S) was placed at the focal plane of the device and was aligned such that only the light focused to one of the two focal points could pass through it. To obtain the reported efficiencies, the optical power passed through the pinhole was measured using a power meter (Thorlabs PM100D with Thorlabs S122C power sensor), and was divided by the power of the incident beam which was measured before the device.  
 

\clearpage
\begin{thebibliography}{10}
\expandafter\ifx\csname url\endcsname\relax
  \def\url#1{\texttt{#1}}\fi
\expandafter\ifx\csname urlprefix\endcsname\relax\def\urlprefix{URL }\fi
\providecommand{\bibinfo}[2]{#2}
\providecommand{\eprint}[2][]{\url{#2}}


\bibitem{Kildishev2013a}
\bibinfo{author}{Kildishev, A.~V.}, \bibinfo{author}{Boltasseva, A.} \&
  \bibinfo{author}{Shalaev, V.~M.}
\newblock \bibinfo{title}{{Planar photonics with metasurfaces.}}
\newblock \emph{\bibinfo{journal}{Science}}
  \textbf{\bibinfo{volume}{339}}, \bibinfo{pages}{1232009}
  (\bibinfo{year}{2013}).

\bibitem{Yu2014}
\bibinfo{author}{Yu, N.} \& \bibinfo{author}{Capasso, F.}
\newblock \bibinfo{title}{{Flat optics with designer metasurfaces}}.
\newblock \emph{\bibinfo{journal}{Nature Mater.}}
  \textbf{\bibinfo{volume}{13}}, \bibinfo{pages}{139--50}
  (\bibinfo{year}{2014}).

\bibitem{Yu2011}
\bibinfo{author}{Yu, N.} \emph{et~al.}
\newblock \bibinfo{title}{{Light propagation with phase discontinuities:
  generalized laws of reflection and refraction}}.
\newblock \emph{\bibinfo{journal}{Science}}
  \textbf{\bibinfo{volume}{334}}, \bibinfo{pages}{333--7}
  (\bibinfo{year}{2011}).

\bibitem{Lin2014a}
\bibinfo{author}{Lin, D.}, \bibinfo{author}{Fan, P.}, \bibinfo{author}{Hasman,
  E.} \& \bibinfo{author}{Brongersma, M.~L.}
\newblock \bibinfo{title}{{Dielectric gradient metasurface optical elements}}.
\newblock \emph{\bibinfo{journal}{Science}} \textbf{\bibinfo{volume}{345}},
  \bibinfo{pages}{298--302} (\bibinfo{year}{2014}).

\bibitem{Lin2013}
\bibinfo{author}{Lin, J.}, \bibinfo{author}{Genevet, P.},
  \bibinfo{author}{Kats, M.~A.}, \bibinfo{author}{Antoniou, N.} \&
  \bibinfo{author}{Capasso, F.}
\newblock \bibinfo{title}{{Nanostructured holograms for broadband manipulation
  of vector beams.}}
\newblock \emph{\bibinfo{journal}{Nano Lett.}} \textbf{\bibinfo{volume}{13}},
  \bibinfo{pages}{4269--74} (\bibinfo{year}{2013}).

\bibitem{Vo2014}
\bibinfo{author}{Vo, S.} \emph{et~al.}
\newblock \bibinfo{title}{{Sub-wavelength grating lenses with a twist}}.
\newblock \emph{\bibinfo{journal}{IEEE Photon. Technol. Lett.}}
  \textbf{\bibinfo{volume}{26}}, \bibinfo{pages}{1375--1378}
  (\bibinfo{year}{2014}).

\bibitem{Monticone2013}
\bibinfo{author}{Monticone, F.}, \bibinfo{author}{Estakhri, N.~M.} \&
  \bibinfo{author}{Al\`{u}, A.}
\newblock \bibinfo{title}{{Full control of nanoscale optical transmission with
  a composite metascreen}}.
\newblock \emph{\bibinfo{journal}{Phys. Rev. Lett.}}
  \textbf{\bibinfo{volume}{110}}, \bibinfo{pages}{203903}
  (\bibinfo{year}{2013}).

\bibitem{Arbabi2014b}
\bibinfo{author}{Arbabi, A.} \& \bibinfo{author}{Faraon, A.}
\newblock \bibinfo{title}{{Fundamental limits of ultrathin metasurfaces}}
  (\bibinfo{year}{2014}).
\newblock arXiv:\eprint{1411.2537}.

\bibitem{Aieta2012}
\bibinfo{author}{Aieta, F.} \emph{et~al.}
\newblock \bibinfo{title}{{Aberration-free ultrathin flat lenses and axicons at
  telecom wavelengths based on plasmonic metasurfaces.}}
\newblock \emph{\bibinfo{journal}{Nano Lett.}} \textbf{\bibinfo{volume}{12}},
  \bibinfo{pages}{4932--6} (\bibinfo{year}{2012}).

\bibitem{Pfeiffer2013}
\bibinfo{author}{Pfeiffer, C.} \& \bibinfo{author}{Grbic, A.}
\newblock \bibinfo{title}{{Cascaded metasurfaces for complete phase and
  polarization control}}.
\newblock \emph{\bibinfo{journal}{Appl. Phys. Lett.}}
  \textbf{\bibinfo{volume}{102}}, \bibinfo{pages}{231116}
  (\bibinfo{year}{2013}).


\bibitem{Fattal2010}
\bibinfo{author}{Fattal, D.}, \bibinfo{author}{Li, J.}, \bibinfo{author}{Peng,
  Z.}, \bibinfo{author}{Fiorentino, M.} \& \bibinfo{author}{Beausoleil, R.~G.}
\newblock \bibinfo{title}{{Flat dielectric grating reflectors with focusing
  abilities}}.
\newblock \emph{\bibinfo{journal}{Nature Photon.}}
  \textbf{\bibinfo{volume}{4}}, \bibinfo{pages}{466--470}
  (\bibinfo{year}{2010}).

\bibitem{Lu2010}
\bibinfo{author}{Lu, F.}, \bibinfo{author}{Sedgwick, F.~G.},
  \bibinfo{author}{Karagodsky, V.}, \bibinfo{author}{Chase, C.} \&
  \bibinfo{author}{Chang-Hasnain, C.~J.}
\newblock \bibinfo{title}{{Planar high-numerical-aperture low-loss focusing
  reflectors and lenses using subwavelength high contrast gratings.}}
\newblock \emph{\bibinfo{journal}{Opt. Express}}
  \textbf{\bibinfo{volume}{18}}, \bibinfo{pages}{12606--14}
  (\bibinfo{year}{2010}).

\bibitem{Klemm2013}
\bibinfo{author}{Klemm, A.~B.} \emph{et~al.}
\newblock \bibinfo{title}{{Experimental high numerical aperture focusing with
  high contrast gratings.}}
\newblock \emph{\bibinfo{journal}{Opt. Lett.}}
  \textbf{\bibinfo{volume}{38}}, \bibinfo{pages}{3410--3}
  (\bibinfo{year}{2013}).

\bibitem{Aieta2015}
\bibinfo{author}{Aieta, F.}, \bibinfo{author}{Kats, M.~A.},
  \bibinfo{author}{Genevet, P.} \& \bibinfo{author}{Capasso, F.}
\newblock \bibinfo{title}{Multiwavelength achromatic metasurfaces by dispersive
  phase compensation}.
\newblock \emph{\bibinfo{journal}{Science}} \textbf{\bibinfo{volume}{347}},
  \bibinfo{pages}{1342--1345} (\bibinfo{year}{2015}).

\bibitem{Warren1995}
\bibinfo{author}{Warren, M.}, \bibinfo{author}{Smith, R.},
  \bibinfo{author}{Vawter, G.} \& \bibinfo{author}{Wendt, J.}
\newblock \bibinfo{title}{High-efficiency subwavelength diffractive optical
  element in gaas for 975 nm}.
\newblock \emph{\bibinfo{journal}{Opt. Lett.}} \textbf{\bibinfo{volume}{20}},
  \bibinfo{pages}{1441--1443} (\bibinfo{year}{1995}).

\bibitem{Lalanne1999}
\bibinfo{author}{Lalanne, P.}, \bibinfo{author}{Astilean, S.},
  \bibinfo{author}{Chavel, P.}, \bibinfo{author}{Cambril, E.} \&
  \bibinfo{author}{Launois, H.}
\newblock \bibinfo{title}{Design and fabrication of blazed binary diffractive
  elements with sampling periods smaller than the structural cutoff}.
\newblock \emph{\bibinfo{journal}{JOSA A}} \textbf{\bibinfo{volume}{16}},
  \bibinfo{pages}{1143--1156} (\bibinfo{year}{1999}).

\bibitem{Arbabi2014}
\bibinfo{author}{Arbabi, A.} \emph{et~al.}
\newblock \bibinfo{title}{{Controlling the phase front of optical fiber beams
  using high contrast metastructures - OSA Technical Digest (online)}}.
\newblock In \emph{\bibinfo{booktitle}{CLEO: 2014}}, \bibinfo{pages}{STu3M.4}
  (\bibinfo{publisher}{Optical Society of America}, \bibinfo{address}{San Jose,
  California}, \bibinfo{year}{2014}).


\bibitem{Arbabi2015}
\bibinfo{author}{Arbabi, A.}, \bibinfo{author}{Horie, Y.},
  \bibinfo{author}{Ball, A.~J.}, \bibinfo{author}{Bagheri, M.} \&
  \bibinfo{author}{Faraon, A.}
\newblock \bibinfo{title}{{Subwavelength-thick lenses with high numerical
  apertures and large efficiency Based on high contrast transmitarrays}}
 \newblock \emph{\bibinfo{journal}{Nature Commun.}}
  \textbf{\bibinfo{volume}{6}}, \bibinfo{pages}{7069} (\bibinfo{year}{2015}).

\bibitem{West2014}
\bibinfo{author}{West, P.~R.} \emph{et~al.}
\newblock \bibinfo{title}{{All-dielectric subwavelength metasurface focusing
  lens}}.
\newblock \emph{\bibinfo{journal}{Opt. Express}}
  \textbf{\bibinfo{volume}{22}}, \bibinfo{pages}{26212} (\bibinfo{year}{2014}).

\bibitem{Decker2015}
\bibinfo{author}{Decker, M.} \emph{et~al.}
\newblock \bibinfo{title}{{High-efficiency dielectric Huygens’ surfaces}}.
\newblock \emph{\bibinfo{journal}{Adv. Opt. Mater.}}
  (\bibinfo{year}{2015}).

\bibitem{Kikuta1997}
\bibinfo{author}{Kikuta, H.}, \bibinfo{author}{Ohira, Y.} \&
  \bibinfo{author}{Iwata, K.}
\newblock \bibinfo{title}{Achromatic quarter-wave plates using the dispersion
  of form birefringence}.
\newblock \emph{\bibinfo{journal}{Appl. Opt.}}
  \textbf{\bibinfo{volume}{36}}, \bibinfo{pages}{1566--1572}
  (\bibinfo{year}{1997}).

\bibitem{Schonbrun2011}
\bibinfo{author}{Schonbrun, E.}, \bibinfo{author}{Seo, K.} \&
  \bibinfo{author}{Crozier, K.~B.}
\newblock \bibinfo{title}{{Reconfigurable imaging systems using elliptical
  nanowires.}}
\newblock \emph{\bibinfo{journal}{Nano Lett.}} \textbf{\bibinfo{volume}{11}},
  \bibinfo{pages}{4299--303} (\bibinfo{year}{2011}).

\bibitem{Yang2014}
\bibinfo{author}{Yang, Y.} \emph{et~al.}
\newblock \bibinfo{title}{{Dielectric meta-reflectarray for broadband linear
  polarization conversion and optical vortex generation.}}
\newblock \emph{\bibinfo{journal}{Nano Lett.}} \textbf{\bibinfo{volume}{14}},
  \bibinfo{pages}{1394--9} (\bibinfo{year}{2014}).

\bibitem{Mutlu2012c}
\bibinfo{author}{Mutlu, M.}, \bibinfo{author}{Akosman, A.~E.},
  \bibinfo{author}{Kurt, G.}, \bibinfo{author}{Gokkavas, M.} \&
  \bibinfo{author}{Ozbay, E.}
\newblock \bibinfo{title}{{Experimental realization of a high-contrast grating
  based broadband quarter-wave plate.}}
\newblock \emph{\bibinfo{journal}{Opt. Express}}
  \textbf{\bibinfo{volume}{20}}, \bibinfo{pages}{27966--73}
  (\bibinfo{year}{2012}).

\bibitem{Zhao2012}
\bibinfo{author}{Zhao, Y.}, \bibinfo{author}{Belkin, M.~A.} \&
  \bibinfo{author}{Al\`{u}, A.}
\newblock \bibinfo{title}{{Twisted optical metamaterials for planarized
  ultrathin broadband circular polarizers.}}
\newblock \emph{\bibinfo{journal}{Nature Commun.}}
  \textbf{\bibinfo{volume}{3}}, \bibinfo{pages}{870} (\bibinfo{year}{2012}).

\bibitem{Garcia-Etxarri2011}
\bibinfo{author}{Garc\'{\i}a-Etxarri, A.} \emph{et~al.}
\newblock \bibinfo{title}{{Strong magnetic response of submicron silicon
  particles in the infrared.}}
\newblock \emph{\bibinfo{journal}{Opt. Express}}
  \textbf{\bibinfo{volume}{19}}, \bibinfo{pages}{4815--26}
  (\bibinfo{year}{2011}).

\bibitem{Evlyukhin2011}
\bibinfo{author}{Evlyukhin, A.~B.}, \bibinfo{author}{Reinhardt, C.} \&
  \bibinfo{author}{Chichkov, B.~N.}
\newblock \bibinfo{title}{{Multipole light scattering by nonspherical
  nanoparticles in the discrete dipole approximation}}.
\newblock \emph{\bibinfo{journal}{Phys. Rev. B}}
  \textbf{\bibinfo{volume}{84}}, \bibinfo{pages}{235429}
  (\bibinfo{year}{2011}).

\bibitem{Spinelli2012}
\bibinfo{author}{Spinelli, P.}, \bibinfo{author}{Verschuuren, M.~A.} \&
  \bibinfo{author}{Polman, A.}
\newblock \bibinfo{title}{{Broadband omnidirectional antireflection coating
  based on subwavelength surface Mie resonators.}}
\newblock \emph{\bibinfo{journal}{Nature Commun.}}
  \textbf{\bibinfo{volume}{3}}, \bibinfo{pages}{692} (\bibinfo{year}{2012}).

\bibitem{Zhan2009}
\bibinfo{author}{Zhan, Q.}
\newblock \bibinfo{title}{{Cylindrical vector beams: from mathematical concepts
  to applications}}.
\newblock \emph{\bibinfo{journal}{Adv. Opt. Photonics}}
  \textbf{\bibinfo{volume}{1}}, \bibinfo{pages}{1} (\bibinfo{year}{2009}).

\bibitem{Phelan2011}
\bibinfo{author}{Phelan, C.~F.}, \bibinfo{author}{Donegan, J.~F.} \&
  \bibinfo{author}{Lunney, J.~G.}
\newblock \bibinfo{title}{{Generation of a radially polarized light beam using
  internal conical diffraction.}}
\newblock \emph{\bibinfo{journal}{Opt. Express}}
  \textbf{\bibinfo{volume}{19}}, \bibinfo{pages}{21793--802}
  (\bibinfo{year}{2011}).

\bibitem{Kozawa2005}
\bibinfo{author}{Kozawa, Y.} \& \bibinfo{author}{Sato, S.}
\newblock \bibinfo{title}{{Generation of a radially polarized laser beam by use
  of a conical Brewster prism}}.
\newblock \emph{\bibinfo{journal}{Opt. Lett.}}
  \textbf{\bibinfo{volume}{30}}, \bibinfo{pages}{3063} (\bibinfo{year}{2005}).

\bibitem{Swanson1989}
\bibinfo{author}{Swanson, G.~J.}
\newblock \bibinfo{title}{{Binary optics technology: the theory and design of
  multi-level diffractive optical elements}}.
\newblock \bibinfo{type}{Tech. Rep.}, \bibinfo{institution}{DTIC Document}
  (\bibinfo{year}{1989}).


\bibitem{Liu2012}
\bibinfo{author}{Liu, V.} \& \bibinfo{author}{Fan, S.}
\newblock \bibinfo{title}{{S4 : A free electromagnetic solver for layered
  periodic structures}}.
\newblock \emph{\bibinfo{journal}{Comput. Phys. Commun.}}
  \textbf{\bibinfo{volume}{183}}, \bibinfo{pages}{2233--2244}
  (\bibinfo{year}{2012}).

\bibitem{Born1999}
\bibinfo{author}{Born, M.} \& \bibinfo{author}{Wolf, E.}
\newblock \emph{\bibinfo{title}{{Principles of optics}}}
  (\bibinfo{publisher}{Cambridge University Press},
  \bibinfo{year}{1999}).

\end{thebibliography}

\clearpage

\textbf{Acknowledgement}\\ This work was supported by Caltech/JPL president and director fund (PDF) and DARPA. Yu Horie was supported as part of the DOE ``Light-Material Interactions in Energy Conversion'' Energy Frontier Research Centre under grand DE-SC0001293 and JASSO fellowship. The device nanofabrication was performed in the Kavli Nanoscience Institute at Caltech.
The authors thank David Fattal and Charles Santori for useful discussion.

\textbf{Author contributions}\\
A. A. and A. F. conceived the experiment. A. A., Y. H., and M. B. fabricated the samples. A. A. performed the simulations, measurements, and analyzed the data. A. A. and A. F. co-wrote the manuscript. All authors discussed the results and commented on the manuscript.

\textbf{Additional information}\\
Correspondence and requests for materials should be addressed to A. F. at faraon@caltech.edu 

\textbf{Competing financial interests}\\ 
The authors declare no competing financial interests.

\newpage

\begin{figure*}[htp]
\centering
\includegraphics[width=5in]{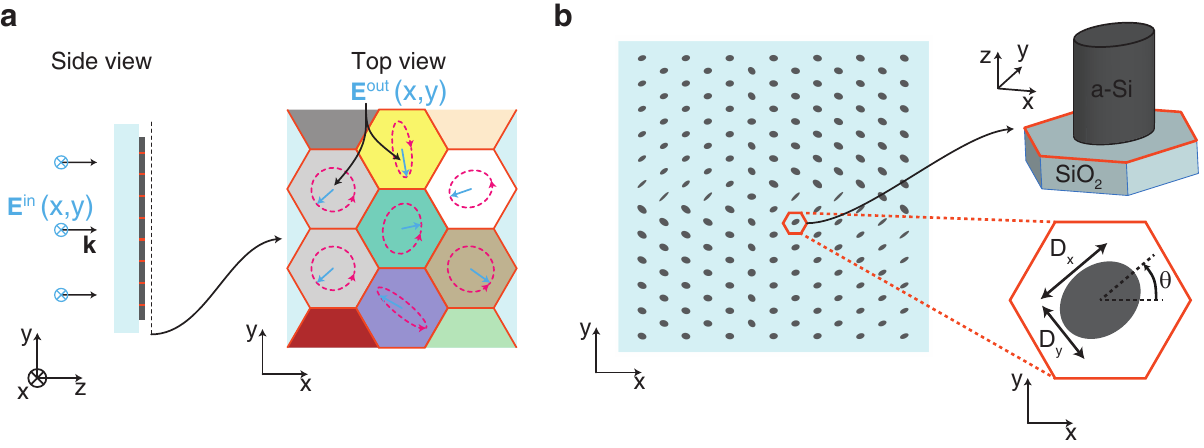}
\caption{\textbf{ Illustration of the proposed metasurface for complete polarization and phase control}. \textbf{a}, Schematic side view (left), and top view (right) of a generic metasurface composed of hexagonal pixels. The polarization and phase of a normally incident optical wave with electric field $\mathbf{E}^\mathrm{in}(x,y)$  is modified at each pixel according to the pixel design. Pixels are coloured differently to emphasize that they can have different designs. In the top view, the spatially varying electric field of the output transmitted light ($\mathbf{E}^\mathrm{out}(x,y)$) at one moment in time, and its polarization ellipse at each pixel are shown by blue arrows and red dashed ellipses, respectively. \textbf{b}, Top view of the proposed implementation of the metasurface. The metasurface is composed of elliptical amorphous silicon posts with the same height, different diameters ($D_x$ and $D_y$), and orientations ($\theta$) which are located at the centres of hexagonal unit cells (pixels). The insets show a schematic 3D view of the amorphous silicon posts and magnified top view. }
\end{figure*}


\begin{figure*}[htp]
\centering
\includegraphics[width=5in]{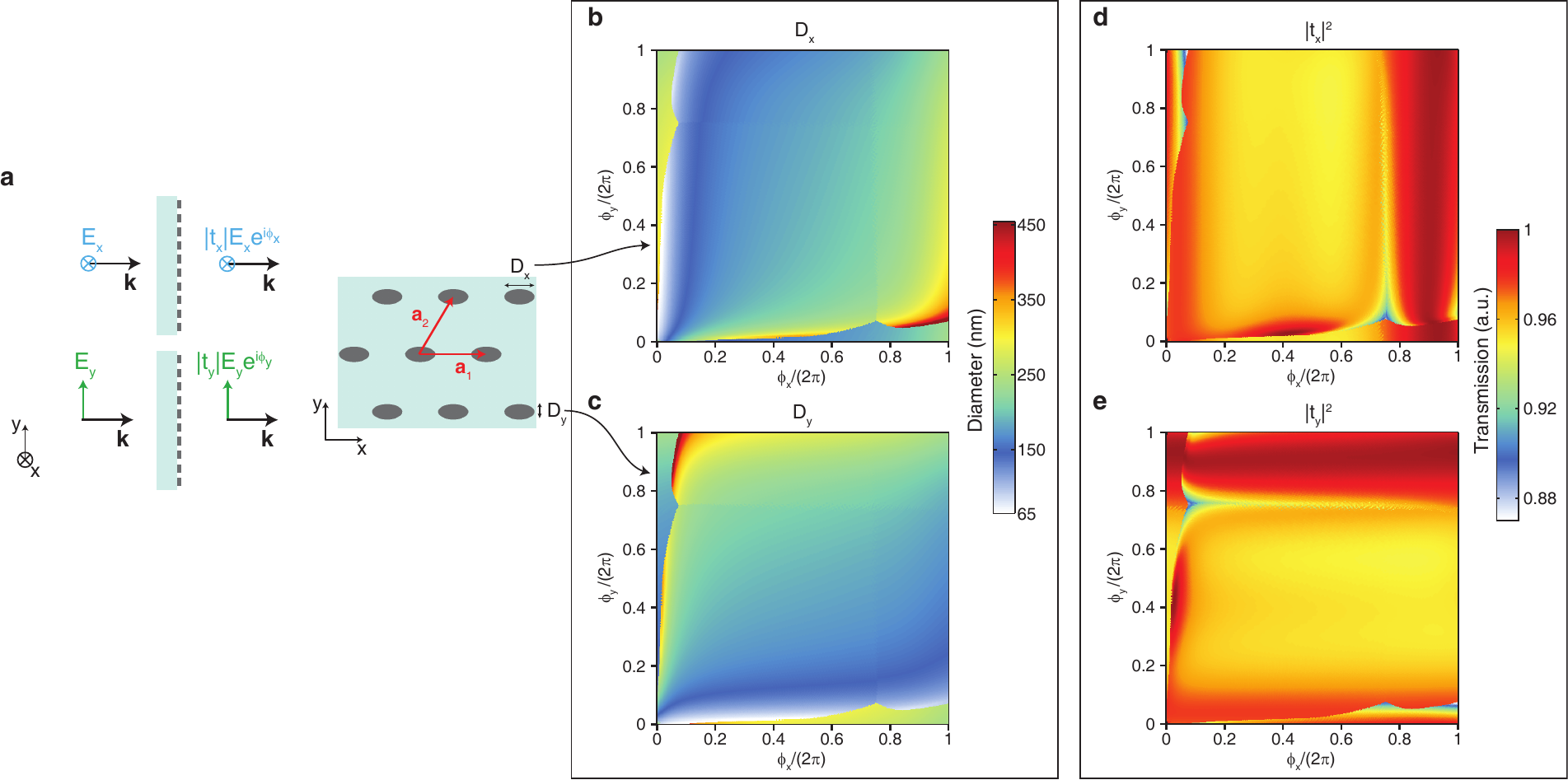}
\caption{\textbf{ Birefringence of the elliptical post arrays.} \textbf{a}, Schematic illustration of a periodic array of elliptical posts with one of the ellipse axes aligned with one of the lattice vectors ($\mathbf{a}_1$). The array exhibits an effective birefringence such that $x$ and $y$-polarized optical waves undergo different phase shifts as they transmit through the array. \textbf{b},\textbf{c}, Simulated colour coded values of the elliptical post diameters ($D_x$ and $D_y$) for achieving $\phi_x$ and $\phi_y$ phase shifts for $x$ and $y$-polarized optical waves, respectively.  To realize a periodic array as shown in (\textbf{a}) which imposes $\phi_x$ and $\phi_y$ phase shifts to $x$ and $y$-polarized optical waves, the diameter of the elliptical posts along $x$ ($D_x$) is obtained from (\textbf{b})  and their corresponding diameter along $y$ ($D_y$) is found from (\textbf{c}).  \textbf{d},\textbf{e}, Simulated colour coded values of the intensity transmission coefficients corresponding to the choice of ellipse diameters shown in (\textbf{b}) and (\textbf{c}). $t_x$ and $t_y$ represent  amplitude transmission coefficients for $x$ and $y$ polarized light, respectively. An operating wavelength of 915~nm, lattice constant of 650~nm, and amorphous silicon post height of 715~nm are assumed (see Methods for the simulation details).}
\end{figure*}


\begin{figure*}[htp]
\centering
\includegraphics[width=5in]{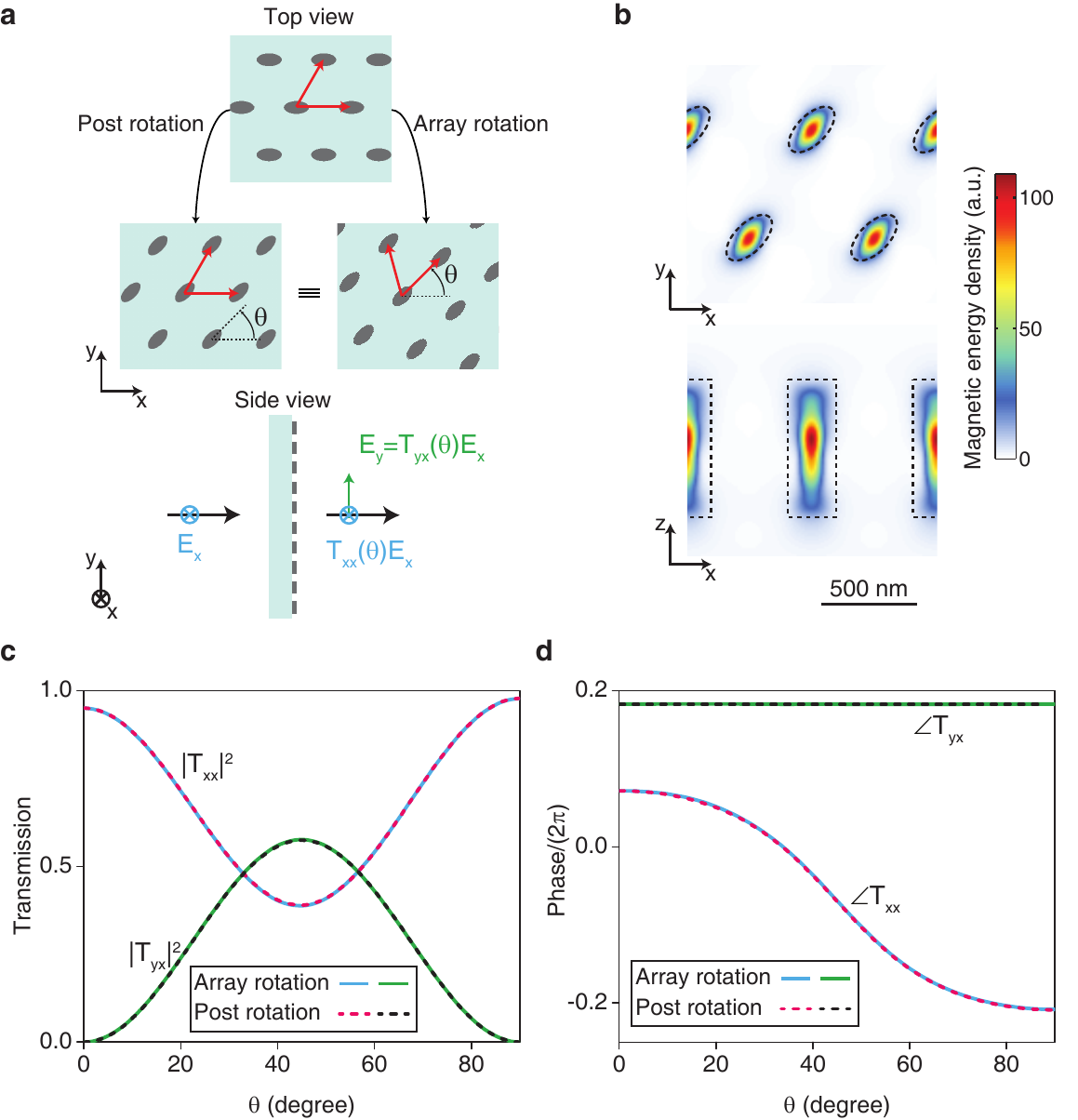}
\caption{\textbf{ Equivalence of the array and post rotations}. \textbf{a}, Schematic illustrations of an array with one of the ellipse diameters aligned to one of the lattice vectors, and two arrays obtained from the first array either by only rotating the elliptical posts, or by rotating the entire array by the same angle $\theta$. Because of birefringence, the rotated arrays convert a portion of the incident $x$-polarized light to $y$-polarized light, as schematically illustrated in the side view. \textbf{b}, Simulated magnetic energy density when light propagates through an array of posts rotated by 45$^\circ$ with respect to the lattice. The dashed black lines depict the boundaries of the posts. $xy$ cross-section on the top, $xz$ cross-section on the bottom. \textbf{c}, \textbf{d}, Simulated values of two elements of the Jones matrices ($T_{xx}$ and $T_{yx}$) of the two arrays shown in (\textbf{a}) as functions of $\theta$. The plots show that the coefficients are almost identical in both cases. See Methods for the simulation details.}
\end{figure*}


\begin{figure*}[htp]
\centering
\includegraphics[width=5in]{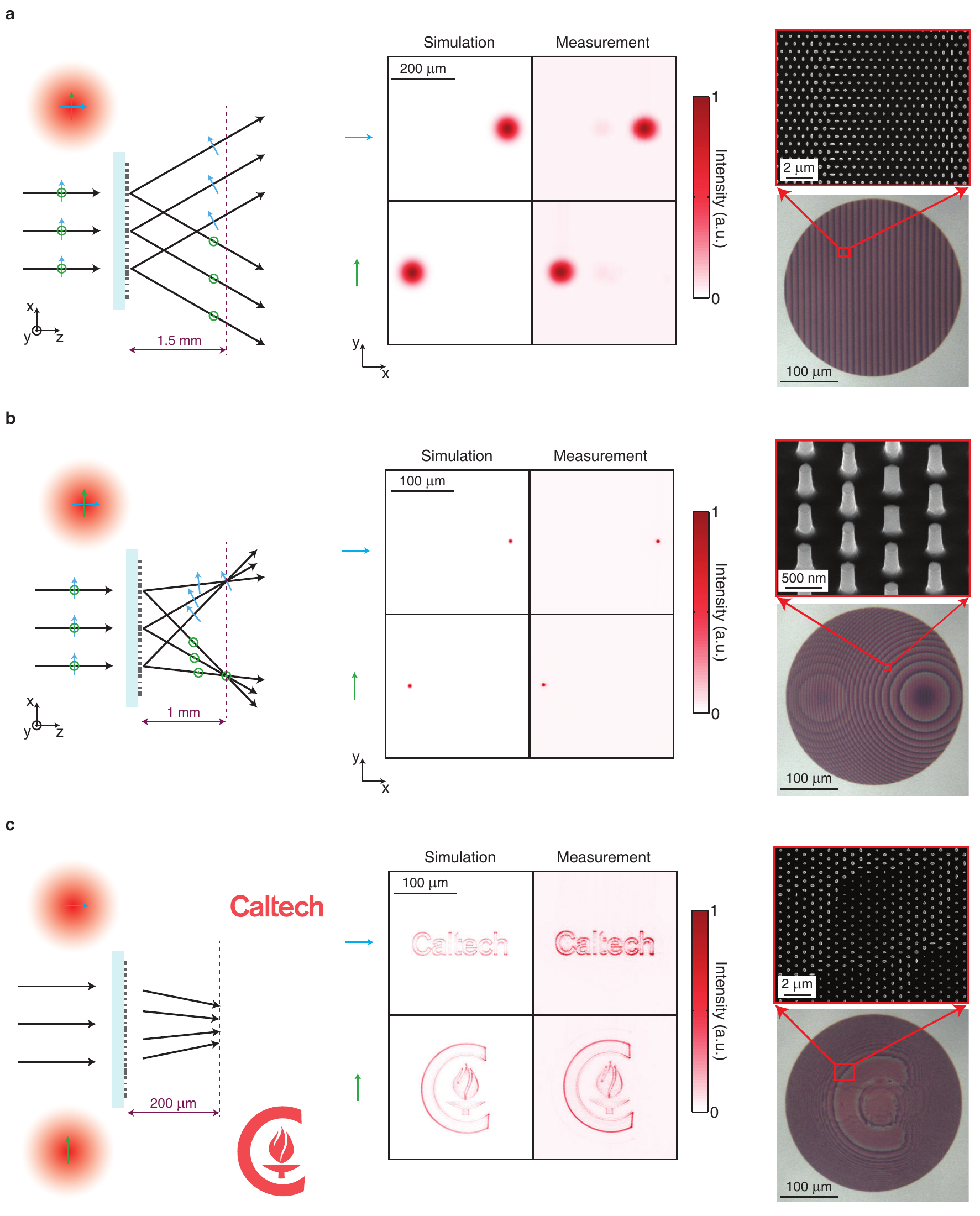}
\caption{\textbf{ Devices for independent control of two polarizations}. \textbf{a}, A polarization beam splitter which separates the $x$ and $y$-polarized light and deflects them by different angles ($\pm 5^\circ$ in this case). The two angles could be chosen at will. \textbf{b}, A device that separates $x$ and $y$-polarized light and focuses them to two different points. The two different points could be chosen at will. \textbf{c}, A polarization switchable phase hologram which generates two arbitrary patterns for $x$ and $y$-polarized light. In this case the word ``Caltech"  is displayed for input $x$ polarization, and an icon is displayed for input $y$ polarization.  Schematic illustration of the devices are depicted on the left, simulated and measured intensity profiles are presented in the middle, and optical (bottom) and scanning electron (top) microscope images of the devices are shown on the right. Blue and green colours correspond to $x$ and $y$ incident polarizations, respectively.}
\end{figure*}


\begin{figure*}[htp]
\centering
\includegraphics[width=5in]{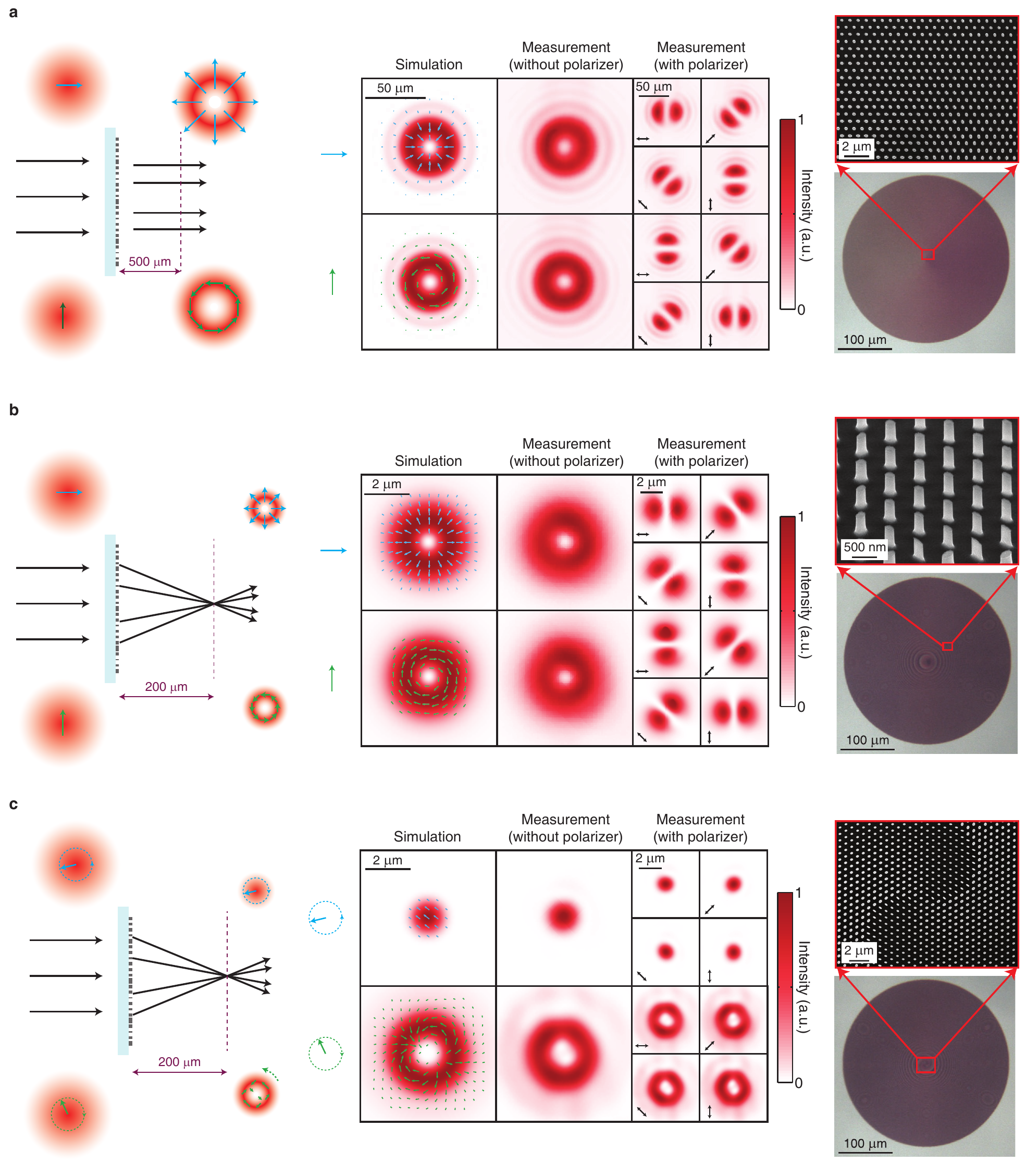}
\caption{\textbf{ Devices for highly efficient vector beam generation}. \textbf{a}, Generation of radially and azimuthally polarized cylindrical vector beams from $x$ and $y$ linearly polarized beams. \textbf{b}, Simultaneous generation and focusing of radially and azimuthally polarized light from $x$ and $y$ linearly polarized beams. \textbf{c}, Device for focal spot control with incident polarization. Incident light is focused to a diffraction limited spot or a doughnut shaped spot depending on its helicity (right-hand or left-hand polarized). Schematic illustration of the devices are depicted on the left, simulated and measured intensity profiles are presented in the middle, and optical (bottom) and scanning electron (top) microscope images of the device are shown on the right. The measured intensity profiles are shown both with and without a linear polarizer placed in front of the camera. In (\textbf{a}) and (\textbf{b}), blue and green colours correspond to $x$ and $y$ incident polarizations, respectively, and in (\textbf{c}) they correspond to right and left handed circularly polarized incident light, respectively. The black double-arrow shows the direction of the polarizer's transmission axis.}
\end{figure*}

\clearpage

\newcommand{\beginsupplement}{%
        \setcounter{table}{0}
        \renewcommand{\thetable}{S\arabic{table}}%
        \setcounter{figure}{0}
        \renewcommand{\thefigure}{S\arabic{figure}}%
     }

\onecolumngrid

      \beginsupplement

\section{Supplementary Information}

\textbf{S1. Arbitrary polarization and phase transformation using symmetric and unitary Jones matrices}

Here we show that any arbitrary polarization and phase transformation can always be performed using a unitary and symmetric Jones matrix. We prove this by determining the unitary and symmetric Jones matrix $\mathbf{T}$ that maps a given input electric field $\mathbf{E}^\mathrm{in}$ to a desired output electric field $\mathbf{E}^\mathrm{out}$.  For polarization and phase transformations (i.e. no amplitude modification), the Jones matrix should be unitary since the transmitted power is equal to the incident power and $|\mathbf{E}^\mathrm{out}|=|\mathbf{E}^\mathrm{in}|$. The general relation between the electric fields of input and output optical waves for normal incidence is expressed  as $\mathbf{E}^\mathrm{out}=\mathbf{T}\mathbf{E}^\mathrm{in}$. For a symmetric and unitary Jones matrix we have
\begin{subequations}
\begin{align}
&T_{xx}E_x^\mathrm{in}+T_{yx}E_y^\mathrm{in}= E_x^\mathrm{out}, \label{eq:Jones_matrix_derivation_1a}\\
&T_{yx}E_x^\mathrm{in}-\frac{T_{yx}}{{T_{yx}}^*}{T_{xx}}^*E_y^\mathrm{in}= E_y^\mathrm{out},
\label{eq:Jones_matrix_derivation_1b}
\end{align}
\end{subequations}
where $E_x^\mathrm{in}$ and $E_y^\mathrm{in}$ are the $x$ and $y$ components of the electric field of the input light, $E_x^\mathrm{out}$ and $E_y^\mathrm{out}$ are the $x$ and $y$ components of the electric field of the output light, $T_{ij}$ ($i,j=x,y$) are the elements of the 2$\times$2 Jones matrix, and * represents complex conjugation.  In deriving Eqs.~\ref{eq:Jones_matrix_derivation_1a} and~\ref{eq:Jones_matrix_derivation_1b}, we have used the symmetric properties $T_{xy}=T_{yx}$, and the unitary condition $T_{xx}{T_{xy}}^*+{T_{yx}}^*T_{yy}=0$. By multiplying Eq.~\ref{eq:Jones_matrix_derivation_1a} by ${T_{xx}}^*$ and Eq.~\ref{eq:Jones_matrix_derivation_1b} by ${T_{yx}}^*$ we obtain 
\begin{subequations}
\begin{align}
&|T_{xx}|^2E_x^\mathrm{in}+T_{yx}{T_{xx}}^*E_y^\mathrm{in}= {T_{xx}}^*E_x^\mathrm{out},\label{eq:Jones_matrix_derivation_2a}\\
&|T_{yx}|^2E_x^\mathrm{in}-T_{yx}{T_{xx}}^*E_y^\mathrm{in}= {T_{yx}}^*E_y^\mathrm{out}.
\label{eq:Jones_matrix_derivation_2b}
\end{align}
\end{subequations}
By adding Eqs.~\ref{eq:Jones_matrix_derivation_2a} and~\ref{eq:Jones_matrix_derivation_2b}, using the unitary condition $|T_{xx}|^2+|T_{yx}|^2=1$, and taking the complex conjugate of the resultant relation, we find
\begin{equation}
T_{xx}{E_x^\mathrm{out}}^*+T_{yx}{E_y^\mathrm{out}}^*={E_x^\mathrm{in}}^*.
\label{eq:Jones_matrix_derivation_3}
\end{equation}
Finally, by expressing Eqs.~\ref{eq:Jones_matrix_derivation_3} and~\ref{eq:Jones_matrix_derivation_1a} in the matrix form, we obtain
\begin{equation}
\begin{bmatrix} {E_x^\mathrm{out}}^* & {E_y^\mathrm{out}}^* \\ E_x^\mathrm{in} & E_y^\mathrm{in} \end{bmatrix}\begin{bmatrix} T_{xx}\\T_{yx} \end{bmatrix}=\begin{bmatrix} {E_x^\mathrm{in}}^*\\E_x^\mathrm{out} \end{bmatrix}.\\ 
\label{eq:Jones_matrix_general_form}
\end{equation}
Therefore, for any given $\mathbf{E}^\mathrm{in}$ and $\mathbf{E}^\mathrm{out}$, we can find $T_{xx}$ and $T_{yx}$ from Eq.~\ref{eq:Jones_matrix_general_form}, and $T_{xy}$ and $T_{yy}$ from the symmetry and unitary conditions as
\begin{subequations}
\begin{align}
T_{xy}&=T_{yx},\\
T_{yy}&=-\mathrm{exp}(2i\angle T_{yx}){T_{xx}}^*.
\end{align}
\label{eq:symmetry_unitary_conditions}
\end{subequations}
Thus, we can always find a unitary and symmetric Jones matrix that transforms any input optical wave $\mathbf{E}^\mathrm{in}$  to any output optical wave $\mathbf{E}^\mathrm{out}$.

\textbf{S2. Realization of any symmetric and unitary Jones matrices using a uniform birefringent metasurface}

Here we show that any symmetric and unitary Jones matrix can be realized using a uniform birefringent metasurface shown in Fig. 2a in the main text, if $\phi_x$, $\phi_y$, and the angle between one of the principal axis of the metasurface and the $x$ axis ($\theta$) could be chosen freely. Any symmetric and unitary matrix is decomposable in terms of its eigenvectors and eigenvalue matrix ($\mathbf{\Delta}$) as
\begin{equation}
\mathbf{T}=\mathbf{V}\begin{bmatrix} \mathrm{e}^{i\phi_x} & 0 \\ 0 & \mathrm{e}^{i\phi_y} \end{bmatrix}\mathbf{V}^T=\mathbf{R}(\theta)\mathbf{\Delta}\mathbf{R}(-\theta),
\label{eq:eigen_decomposition}
\end{equation}
where superscript \textit{T} represents the matrix transpose operation. $\mathbf{V}$ is a real unitary matrix; therefore, it corresponds to an in-plane geometrical rotation $\mathbf{R}$ by an angle that we refer to as $\theta$, and since $\mathbf{V}^T=\mathbf{V}^{-1}$, $\mathbf{V}^T$ represents a rotation by $-\theta$. According to Eq.~\ref{eq:eigen_decomposition}, the operation of a metasurface that realizes the Jones matrix $\mathbf{T}$ can be considered as rotating the electric field of the input wave ($\mathbf{E}^\mathrm{in}$) by $-\theta$, phase shifting the $x$ and $y$ components of the rotated $\mathbf{E}^\mathrm{in}$ respectively by $\phi_x$ and $\phi_y$, and rotating back the rotated and phase shifted vector by angle $\theta$. Equivalently, $\mathbf{T}$ can be implemented using a metasurface that imposes phase shifts $\phi_x$ and $\phi_y$ to the components of $\mathbf{E}^\mathrm{in}$ along angles $\theta$ and $90^{\circ}\!+\!\theta$, respectively. Such a metasurface is realized by starting with a metasurface whose principal axis are along $x$ and $y$ directions and imparts $\phi_x$ and $\phi_y$ phase shifts to $x$ and $y$-polarized waves, and rotating it anticlockwise by angle $\theta$. Therefore, any symmetric and unitary Jones matrix can be realized using a metasurface if its $\phi_x$, $\phi_y$, and in-plane rotation angle ($\theta$) could be chosen freely.

\textbf{S3. Independent wavefront control for two orthogonal polarizations}

In this section, we derive the necessary condition for the design of a device that imposes two independent phase profiles to two optical waves with orthogonal polarizations. The four elements of the Jones matrix $\mathbf{T}$ are found uniquely using Eqs.~\ref{eq:Jones_matrix_general_form} and~\ref{eq:symmetry_unitary_conditions}, when the determinant of the matrix on the left hand side of Eq.~\ref{eq:Jones_matrix_general_form} is nonzero. Therefore,  a devices that is designed to map $\mathbf{E}^\mathrm{in}$ to $\mathbf{E}^\mathrm{out}$, converts an optical wave whose polarization is orthogonal to $\mathbf{E}^\mathrm{in}$  to an optical wave polarized orthogonal to $\mathbf{E}^\mathrm{out}$. For example, an optical element designed to generate radially polarized light from $x$ polarized input light, will also generate azimuthally polarized light from $y$ polarized input light.   

In the special case that the determinant of the matrix on the left had side of Eq.~\ref{eq:Jones_matrix_general_form} is zero we have
\begin{equation}
{E_x^\mathrm{out}}^*E_y^\mathrm{in}-{E_y^\mathrm{out}}^*E_x^\mathrm{in}=0,
\end{equation}
and because $\mathbf{T}$ is unitary we have $|\mathbf{E}^\mathrm{in}|=|\mathbf{E}^\mathrm{out}|$; therefore we find $\mathbf{E}^\mathrm{out}=\exp(i\phi){\mathbf{E}^\mathrm{in}}^*$ where $\phi$ is an arbitrary phase. This special case corresponds to a device that preserves the polarization ellipse of the input light, switches its handedness (helicity), and imposes a phase shift on it. In this case, the $\mathbf{T}$ matrix is not uniquely determined from Eq.~\ref{eq:Jones_matrix_general_form}, and an additional condition, such as the phase profile for the orthogonal polarization, can be imposed on the operation of the device. Therefore, the device can be designed to realize two different phase profiles for two orthogonal input polarizations.

\clearpage
\section{Supplementary Figures}
\begin{figure*}[htp]
\centering
\includegraphics[width=\columnwidth]{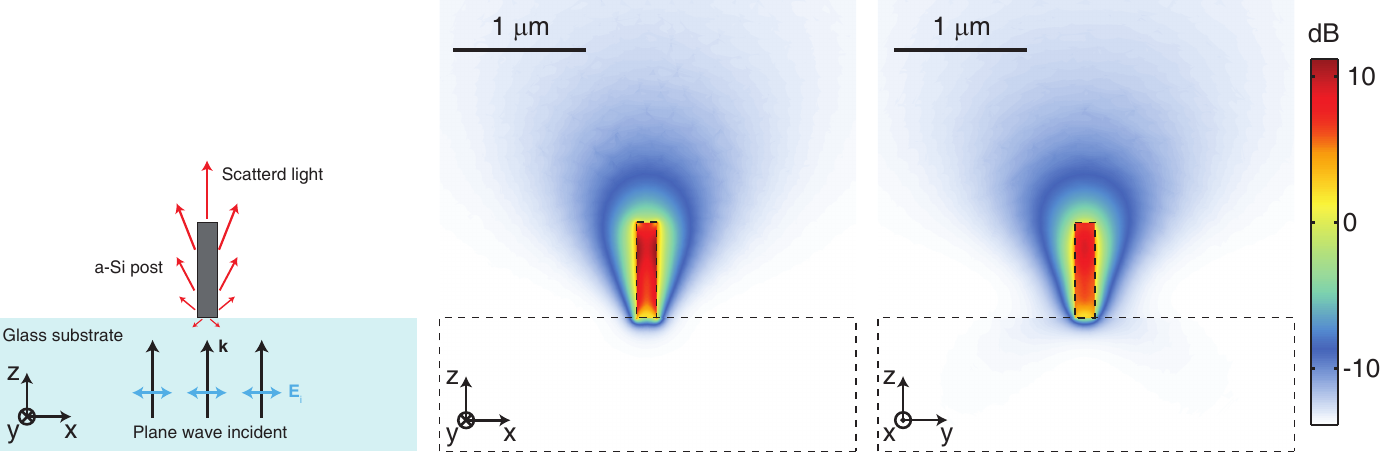}
\caption{\textbf{Large forward scattering by a single amorphous silicon post}. Schematic illustration and finite element simulation results of light scattering by a single 715 nm tall circular amorphous silicon post with a diameter of 150 nm. The simulation results show the logarithmic scale energy density of the light scattered by the single amorphous silicon post over the $xz$ and $yz$ planes. The energy densities are normalized to the energy density of the 915 nm $x$-polarized incident plane wave.}
\end{figure*}

\clearpage
\begin{figure*}[htp]
\centering
\includegraphics[width=0.85\columnwidth]{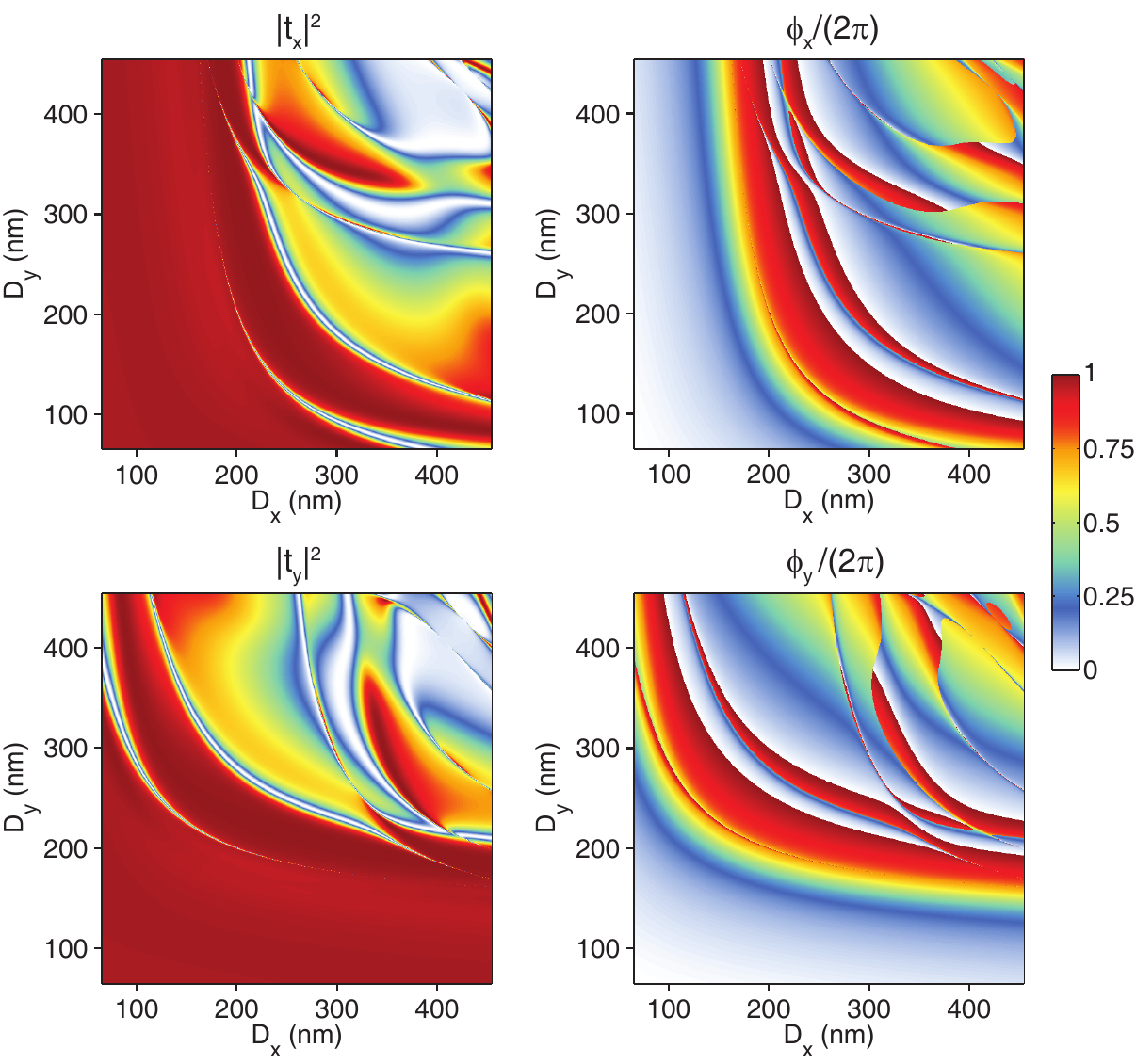}
\caption{\textbf{Phase shifts and intensity transmission coefficients as a function of elliptical post diameters, used to derive data in Fig. 2b-e of the main text}. Intensity transmission coefficients ($|t_x|^2$ and $|t_y|^2$) and the phase of transmission coefficients ($\phi_x$ and $\phi_y$) of $x$ and $y$-polarized optical waves for the periodic array of elliptical posts shown in Fig. 2a of the main text as functions of the post diameters.}
\end{figure*}

\clearpage
\begin{figure*}[htp]
\centering
\includegraphics[width=0.9\columnwidth]{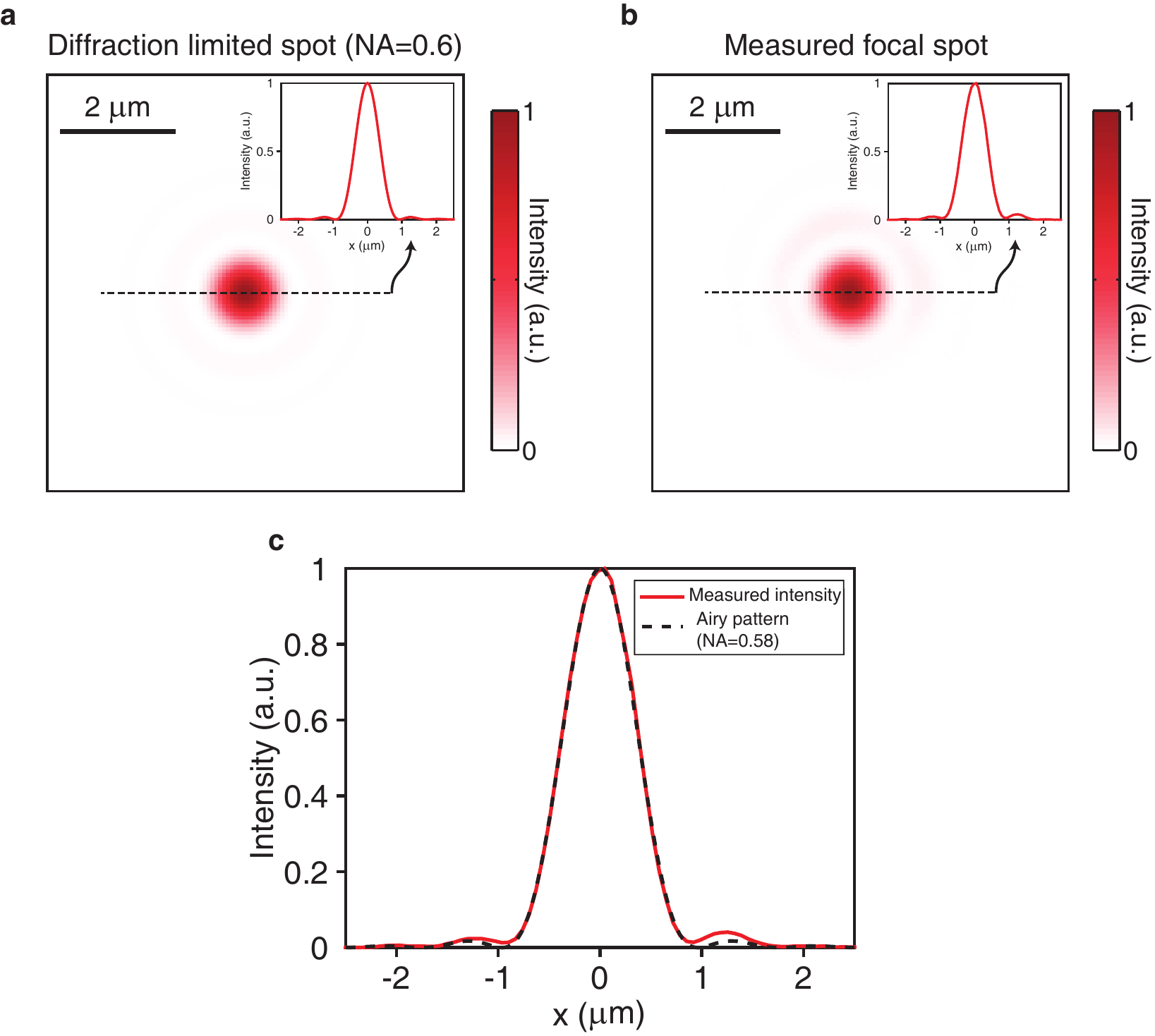}
\caption{\textbf{Diffraction limited focusing by device shown in Fig. 5c}. \textbf{a}, Theoretical diffraction limited focal spot (Airy disk) for a lens with numerical aperture (NA) of 0.6 at the operation wavelength of 915 nm. Inset shows the intensity along the dashed line.\textbf{b}, Measured focal spot for the device shown in Fig. 5c when the device is uniformly illuminated with right handed circularly polarized 915 nm light. Inset shows the intensity along the dashed line. \textbf{c}, Measured intensity along the dashed line shown in (\textbf{b}) and its least squares Airy pattern fit which has an NA of 0.58.}
\end{figure*}

\clearpage
\begin{figure*}[htp]
\centering
\includegraphics[width=0.97\columnwidth]{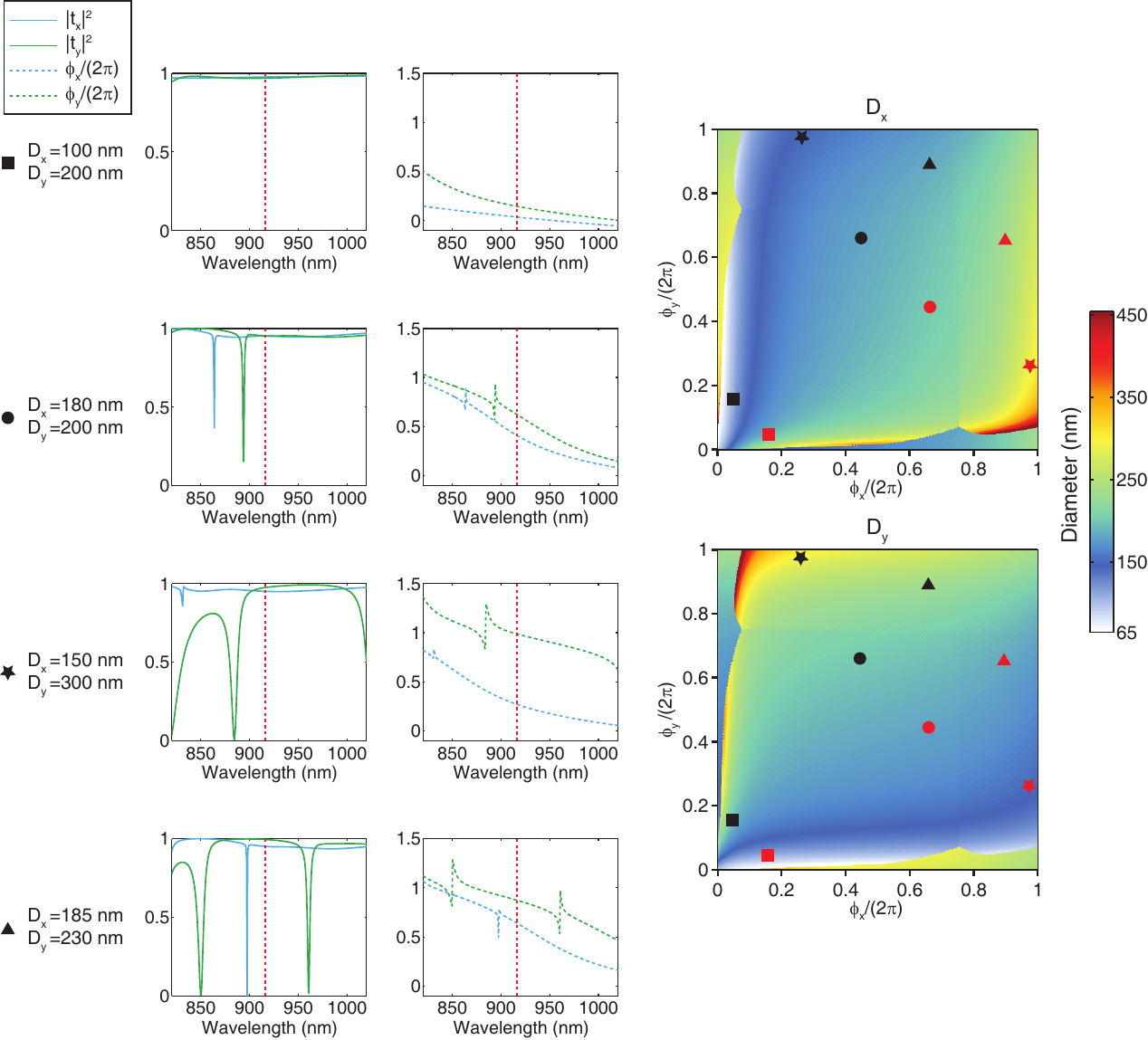}
\caption{\textbf{Transmission spectra of periodic arrays of elliptical posts showing that the operation wavelength does not overlap with resonances}. 
Wavelength dependence of the intensity transmission coefficients ($|t_x|^2$ and $|t_y|^2$) and the phase of transmission coefficients ($\phi_x$ and $\phi_y$) of  $x$ and $y$-polarized optical waves for the periodic arrays schematically shown in Fig. 2a of the main text. The spectra are shown for a few arrays with different ($D_x$, $D_y$) combinations: (100 nm, 200 nm), (180 nm, 200 nm), (150 nm, 300 nm), (185 nm, 230 nm). The corresponding phase shift values and post diameters for these arrays are shown on the $D_x$ and $D_y$ graphs on the right with black symbols. For brevity, only the spectra for the arrays with $D_y>D_x$ are shown. The transmission and phase spectra for the arrays with $D_x>D_y$ (which are shown with red symbols on the $D_x$ and $D_y$ graphs) can be obtained by swapping $x$ and $y$ in the spectra graphs.  The desired operation wavelength ($\lambda=$915 nm) is shown with dashed red vertical lines in the spectra plots, and it does not overlap with any of the resonances of the periodic arrays.}
\end{figure*}

\clearpage
\begin{figure*}[htp]
\centering
\includegraphics[width=\columnwidth]{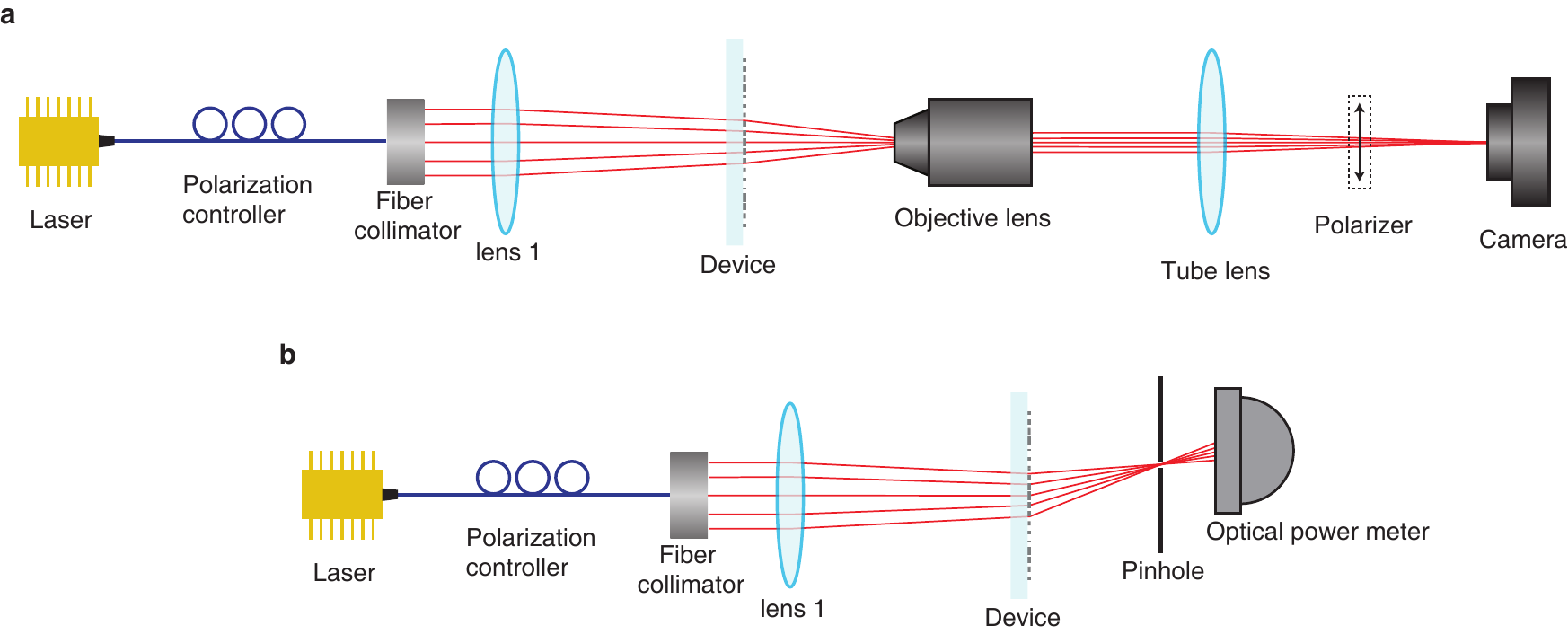}
\caption{\textbf{Measurement setup}. \textbf{a}, Schematic illustration of the measurement setup used for characterization of devices modifying polarization and phase of light. The linear polarizer was inserted into the setup only during the polarization measurements. \textbf{b}, Schematic drawing of the experimental setup used for efficiency characterization of the device shown in Fig. 4b.}
\end{figure*}

\end{document}